# Data-Driven EV Charging Load Profile Estimation and Typical EV Daily Load Dataset Generation

Linhan Fang, Jesus Silva-Rodriguez, Xingpeng Li
Department of Electrical and Computer Engineering
University of Houston
Houston, TX, USA

*Abstract*—Widespread electric vehicle (EV) adoption introduces new challenges for distribution grids due to large, localized load increases, stochastic charging behavior, and limited data availability. This paper proposes two data-driven methods to estimate residential EV charging profiles using real-world customer meter data from CenterPoint Energy serving the Houston area. The first approach applies a least-squares estimation to extract average charging rates by comparing aggregated EV and non-EV meter data, enabling a statistical method for starting and ending charge times. The second method isolates EV load from meter profiles and applies a kernel density estimation (KDE) to develop a probabilistic charging model. Both methods produce a distinct "u-shaped" daily charging profile, with most charging occurring overnight. The validated profiles offer a scalable tool for utilities to better anticipate EV-driven demand increases and support proactive grid planning.

## I. INTRODUCTION

The transportation sector accounts for approximately 62.3% of global fuel consumption and contributes to a quarter of global greenhouse gas emissions [1]. The rapid development and widespread adoption of electric vehicles (EV) are key strategies for mitigating these impacts, offering significant environmental benefits while reducing fossil fuel dependency [2]. The International Energy Agency (IEA) reported that the global EV fleet reached 58 million units by the end of 2024, with 4 more million units sold in the first quarter of 2025, and projection of these numbers to quadruple by 2030 [3]. However, this large-scale integration presents substantial challenges to the stability and security of power distribution networks. Uncoordinated EV charging load can exacerbate peak loads and cause severe instability issues [4]-[5].

Accurately modeling EV charging load is essential for grid planning and operation, but it remains a complex task. The spatiotemporal distribution of charging demand is highly dependent on stochastic user behaviors, including route selection and charging preferences [6]-[7]. This modeling challenge is compounded by data limitations. Due to the inherent variability of charging demand and user privacy concerns, previous studies have often relied on data from limited charger experiments of small-scale, actual consumption records [8]. To improve the fidelity of EV charging load modeling, more reasonable and effective data extraction methodologies as well as more accurate daily EV load estimations are required.

The majority of data-driven EV charging demand forecasting studies use historical EV charging demand data. Unfortunately, privacy concerns often limit access to this data, and as a result, there is a very limited number of publicly available EV charging datasets [9]. This limitation hinders accurate forecasting, since some of the more sophisticated forecasting models, such as deep learning models, depend heavily on the training data utilized. Moreover, many of the current algorithms for EV charging load forecasting are based on the assumption that charging events and their durations are mostly determined by the state-of-charge (SOC) of the EV's battery at the time of arrival due to the strong correlation often found between EV charging load and this parameter [10]. Therefore, exploring starting charge times, ending charge times, charging event duration, and SOC data can lead to the most accurate prediction strategies.

This paper presents two methods to generate daily EV charging profile estimations based on real data from CenterPoint Energy's (CNP) service areas in the greater Houston area. The methods are based on estimated starting and ending charging times and charging durations determined from the noticeable differences that are observed from load data from consumer meters with and without level-2 EV chargers registered. The resulting charging profiles are validated by comparison with other publicly available residential profile data that show very similar daily charging behavior for EV users. Moreover, these estimated charging profiles, as well as the synthesized data generated from the meter load data from CNP, are meant to present additional case scenarios for daily EV charging patterns in the Texas gulf coast region.

## II. STATISTICAL AGGREGATE CONSTANT LOAD METHOD

This method consists of the following: using peak load data for single-phase laterals serving a series of customers, the average EV charging load is determined based on the total number of customers with and without level-2 EV chargers registered, and their average peak load. Once average charging load is determined, then average daily load profiles for non-EV and EV meters are obtained from two years of individual meter data, and an average daily load difference between EV and non-EV meters is derived, from which starting and ending charge times can be estimated. Different probability distributions can be assumed using statistics from this load difference, from which starting and ending charge times can be sampled, and the charging rate is assumed to be constant at the level determined based on the peak load data. The different steps of this method are explained in the following subsections.

### A. EV Charging Rate State Estimation

An estimate of the average EV charging load is derived based on the peak load data for different line fuses of the single-

---

phase laterals for a CNP service area. For an $N$ number of line fuses, a linear system of $N$ equations of only two variables can be adopted. The equations will have the following form:

$$a_n X + b_n Y = P_n^{peak}, \quad (1)$$

where $a_n$ and $b_n$ are the number of non-EV meters and EV meters at line fuse $n$, respectively, $X$ and $Y$ are the average power associated to non-EV and EV meters, respectively, and $P_n^{peak}$ is the peak power recorded for line fuse $n$.

Since solving a system of two variables having more than two independent equations is infeasible, then a state estimation method is carried out based on a least-squares approach. For two state variables when having more than two measurements, then an estimate can be obtained for what those two variables should be [11]. This idea consists of minimizing the following:

$$\min J(X, Y) = \sum_{n=1}^{N} \frac{\left(P_n^{peak} - a_n X + b_n Y\right)^2}{\sigma_n^2}, \quad (2)$$

where $\sigma_n$ is the standard deviation of the peak power recorded at line fuse $n$.

In matrix form, $J$ can be written as

$$J(\theta) = [P^{peak} - H\theta]^T R^{-1} [P^{peak} - H\theta], \quad (3a)$$

with,

$$\theta = \begin{bmatrix} X \\ Y \end{bmatrix}, \quad (3b)$$

$$H = [a \ b], \quad (3c)$$

$$R = \sigma I, \quad (3d)$$

where $I$ is an $N*N$ identity matrix. Now, to minimize $J$, its gradient must be obtained and be set equal to zero to later solve for the estimated values of $X$ and $Y$. The gradient of $J$ is given by

$$\nabla_\theta J(\theta) = -2H^T R^{-1} P^{peak} + 2H^T R^{-1} P^{peak} H, \quad (4)$$

and solving for $\theta$ then gives the following:

$$\theta = \begin{bmatrix} X \\ Y \end{bmatrix} = [H^T R^{-1} H]^{-1} H^T R^{-1} P^{peak}. \quad (5)$$

There are line fuse data for total number of meters, number of EV meters, and peak load downstream of each line fuse, for 72 different line fuses in a CNP service area. Using the data for peak load recorded for each line fuse, the estimated average power for EV meters, non-EV meters, and for the average EV charging power are listed in Table II. Therefore, the average EV charger in this service area can be expected to be rated at 10.5 kW. This EV charging load is consistent with expected EV charging rates for level-2 chargers, which can range from 7 to 19 kW [12].

TABLE I: Average load determined with state estimation.

| Non-EV Meter Avg. Peak Power ($X$) [kW] | EV Meter Avg. Peak Power ($Y$) [kW] | Avg. EV Charging Power ($Y$-$X$) [kW] |
|---|---|---|
| 6.4567 | 16.9860 | 10.5294 |

Assuming possible maximum and minimum rated power for level-2 EV chargers of 7 and 19 kW, respectively, and a mean of 10.53 kW, the rated power of each EV charger at each EV meter in the service area can be randomly sampled from a truncated exponential distribution constructed from these three parameters [13]-[14]. The PDF for this distribution is shown in Fig 1, which can estimate rated power for each EV between 7 and 19 kW, but centered around 10.53 kW.

B. *Daily Average Consumer Load*

For 7 out of the 72 line fuses had individual meter load data available. Each with their respective number of total meters as well as how many of those have EV chargers registered. The individual meter load datasets for these line fuses present complete yearly data for the years 2022 and 2023 in 15-minute intervals. Taking the average daily load profile for all the days of these two years for non-EV and EV meters demonstrates that there is a clear difference between the load profile shapes of these two, with the EV meters having higher load peaks in the late evening and night, and comparable load only between morning and afternoon hours. This can be seen in the plot for non-EV and EV average daily load profiles in Fig. 1 and the plot for the average load difference between EV and non-EV meters in Fig. 2.

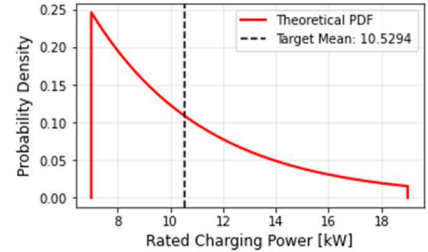

Fig. 1: Truncated exponential distribution PDF for rated power of EV chargers.

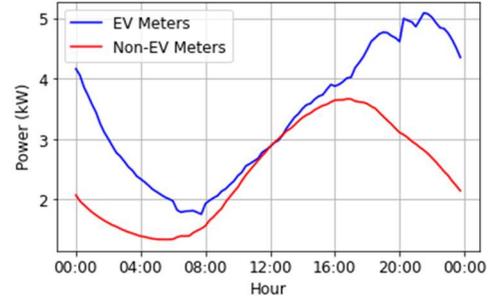

Fig. 2: Average daily load profiles for EV and non-EV meters.

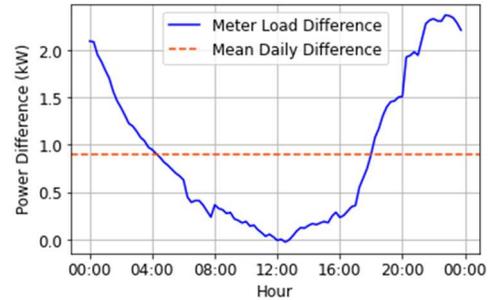

Fig. 3: Average load difference between EV and non-EV meters.

C. *Starting and Ending Charge Time Probabilities*

Statistical methods can then be used to estimate when EVs begin and end their charging sessions based on the average load difference between EV and non-EV meters. For example, the average starting charge time can be taken to be at 18:00 (6 PM), which is when the load difference crosses the daily average load difference of Fig. 2, and for the average ending charge time to be around 04:15 (4:15 AM). Consequently, the standard deviation for both starting and ending charge times can be assumed as the time difference between these averages starting and ending charge times, and the closest time interval at which the load difference is zero. This statistical information for starting and ending charge times is summarized in Table I.

TABLE II: Statistical parameters for normal distributions used to estimate starting and ending EV charge times.

| Starting Charge Time | | Ending Charge Time | |
|---|---|---|---|
| Mean (μ) | Standard Deviation (σ) | Mean (μ) | Standard Deviation (σ) |
| 18:00 | 1.92 hrs | 04:15 | 2.58 hrs |

Using this statistical information, different probability distributions can be derived to estimate individual probabilities for a single EV to begin and end charging around these averages. The probability distributions chosen to model the starting and ending charge time probabilities based on the statistical parameters of Table I are a log-normal distribution and a gamma distribution [15], whose curves for starting and ending charge times are given in Fig. 4.

Moreover, the ending charge time can also be understood to be when the EV is disconnected from the charger. Which means that the EV may be fully charged before that time is reached. However, without access to data on initial SOC at the start of every charging session, the best estimate can be to assume that, if the overall charge duration results in more energy drawn than the typical capacity of an EV (i.e., 60 kW [12]), the ending charge time may be truncated once that capacity is reached. This will be shown in the resulting EV profile estimates using this method, presented in Section IV.

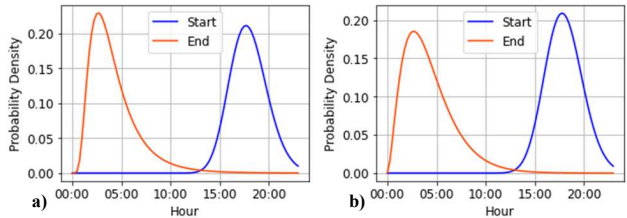

Fig. 4: a) Log-normal distribution, and b) gamma distribution for starting and ending charge times.

### III. NORMALIZED SUBTRACTION METHOD

The second proposed method is a multi-stage process designed to extract the EV charging load from total household electricity consumption data. The process commences with the generation of a reference baseline signal for non-EV users, which is created using averaging and normalization techniques. Concurrently, the aggregated EV user profile is adjusted to the same magnitude as this reference baseline. A critical step for accurate load isolation is temporal alignment; this is accomplished by synchronizing the 95th percentile valley points of the two signals. Following alignment, the non-EV baseline is subtracted from the amplitude-matched EV user signal, yielding the charging component. The procedure concludes with an amplitude restoration step, applying an inverse transformation to return the extracted charging signal to its original magnitude.

#### A. Extracting EV Charging Curves

This method uses the same electricity consumption dataset as the first method of Section II, also dividing it into electricity consumption data for EV users and electricity consumption data for non-EV users. In addition, the fact that the most prominent difference between the load demand of EV and non-EV users is mostly an increased amplitude in late evening hours is emphasized and considered in this method. Therefore, the electricity consumption data of all non-EV users under the same line fuse are averaged and then normalized to obtain a baseline curve under each line fuse, which can represent the electricity consumption pattern of non-EV users. Then, the scaling factors are calculated according to the variation range of the baseline curve and the variation range of the EV users' electricity consumption data, so that they maintain the same fluctuation trend but different absolute amplitudes. Fig. 5 shows a three-day data comparison between one EV meter curve and the non-EV meters baseline curve. The portion of the statistics where EV meter data clearly violates non-EV meters electricity consumption behavior is categorized as EV charging load.

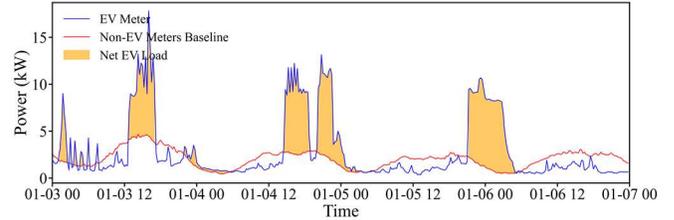

Fig. 5: Synchronization of non-EV baseline signal and EV meter signal.

#### B. EV Charging Curve Probability Distribution Function

The EV charging data is analyzed based on the extracted charging events data, and specific indicators include the charging energy, average power and the probability distribution of the start and end times of the charging events. These features are typically statistically summarized and plotted as histograms, then fitted to various probability distribution functions (PDFs) for subsequent analysis. However, since the indicators obtained by this method are difficult to fit using standard analytical distribution functions, kernel density estimation (KDE) is used to construct a data-driven probabilistic model which is a nonparametric technique used to estimate PDF of continuous random variables. It uses a density estimate by superimposing kernel functions centered at each observation data point. This yields a smooth, data-driven approximation of the empirical distribution. In the probability distribution model for charging energy, average power, charging start time and end time, each horizontal axis value corresponds to a probability value.

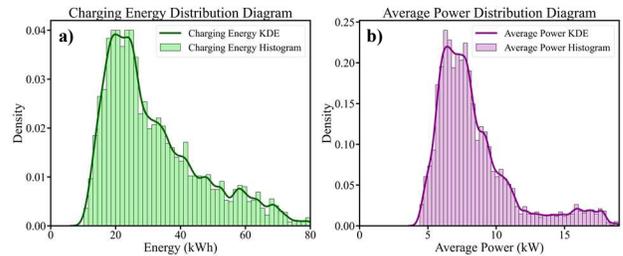

Fig. 6: a) Charging energy and b) average charging power distribution diagrams.

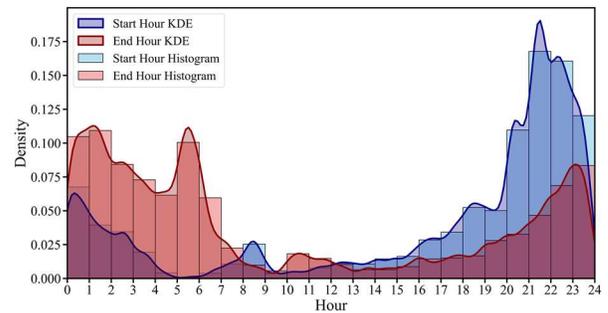

Fig. 7: Charging start time and end time distribution diagram.

As shown in Fig. 6 and 7, the most charging events involve less than 50 kWh, peaking at approximately 20 kWh and average charging power ranges from about 5 kW to 18 kW. Charging start times are mainly concentrated from 19:00 to 24:00, with end times mainly occurring between 23:00 and 5:00. Furthermore, the distribution of charging events in the two methods presented in this paper is similar and comparable.

## IV. EV Charing Load Profile Estimations

### A. Statistical Aggregate Constant Load Method

As an example, a service area with 144 EVs is assumed, where every single EV can be assigned a random starting and ending charge time sampled from the distributions of Fig. 4, and a random rated EV charger power sampled from Fig. 1. This generates collective charging load profiles that take the forms shown in Fig. 8 for a log-normal and a gamma distribution.

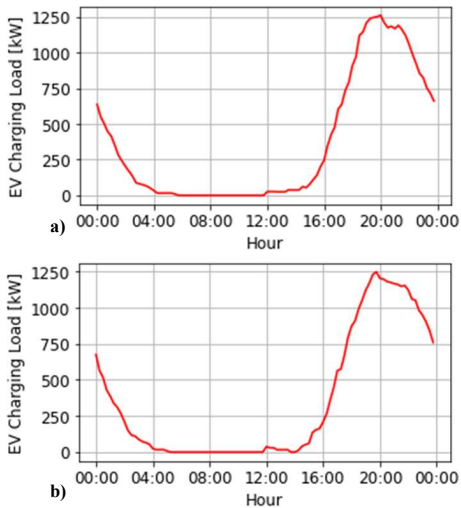

Fig. 8: Estimated EV charging load profile for 144 EVs using a) a log-normal distribution, and b) a gamma distribution.

It is important to highlight that these estimates assume that when EVs are charging they are always consuming constant power, which may not necessarily be the case, as it is known that level-2 charges have a varying charging rate that can go from 7 kW to 19 kW, depending on SOC [9]-[10], [12]; however, SOC of each EV in CNPs service area is not captured in the dataset, as this is customer's private information not reported to the electric utility. Therefore, for a more precise daily profile, data on SOC at the beginning of the charging session as well as at the end of the charging session is also needed. Nonetheless, these daily EV charging load profile estimates can serve as a good planning resource for estimating potential load increases in a service area at different levels of EV adoptions.

### B. Normalized Subtraction Method

To ensure charging event fidelity, we utilized a joint probability model for power and energy demand. This model defines a two-phase charging profile based on sampled values: for demands exceeding 50 kWh, the sampled average power is applied for the first 50 kWh, after which the power is reduced to 70% of that average. For demands of 50 kWh or less, the average power is applied for the initial ninety percent of the energy transfer, followed by a reduction to 70% of the average for the final 10%. This methodology simulates the characteristic power taper-off associated with a high SOC. Utilizing this model, a Monte Carlo simulation is conducted to generate 144 annual charging scenarios, assuming a homogeneous EV fleet and a 90% daily charging probability.

Fig. 9 presents a 24-hour average load profile for a simulated cohort of 144 EVs, generated using the KDE model. The ordinate quantifies the aggregate power demand in kilowatts, while the abscissa represents the time of day. The visualization comprises two key components: a central solid line indicating the mean power demand, and a surrounding shaded region that encapsulates the load's variability bandwidth, likely representing the statistical range of the simulated demand.

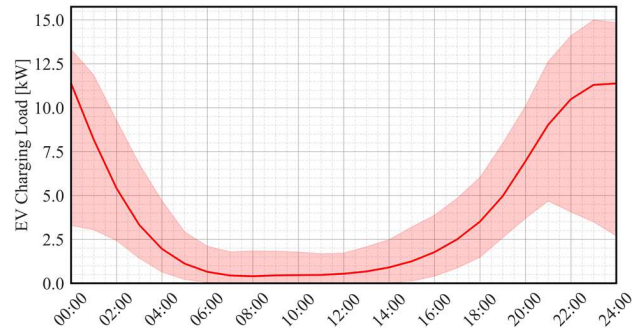

Fig. 9: Generated daily EV charging load profile for 144 EVs using the KDE distribution.

Quantitatively, the load profile exhibits a distinct U-shaped topology. A sustained load nadir is observed during diurnal hours, specifically between approximately 08:00 and 16:00, where the mean demand stabilizes at a minimum of approximately 1.2 kW. Conversely, a significant ramp-up phase initiates post-16:00, with the load ascending to a prolonged plateau during nocturnal hours. During this peak period, the mean aggregate power demand almost exceeds 11 kW. It is also observed that the variability bandwidth is non-uniform, and that it is minimized during the off-peak nadir and expands to its maximum width concurrent with the evening load peak. Critically, the stochasticity of the load is most pronounced during this peak charging window. This indicates that while charging events are temporally clustered, significant heterogeneity persists in individual vehicle charging start times, durations, and power levels. In addition, based on different charging rates, we divide the EV into four categories: 4-7 kW, 7-11.2 kW, 11.2-15 kW, and above 15 kW, and generate annual charging data.

## V. Public Datasets Comparisons

This section presents the datasets from 2 publicly available sources for residential charging, which serve as a comparison with the trends obtained by the two methods of this paper. The first source corresponds to residential EV charging in the UK, which reported the most popular time for EV plug-in was between 17:00 and 18:59, and the most popular time for EV plug-out was between 07:00 and 09:00 [16]. This data is from 2017, and it is recognized that the majority of users charged their EVs with level 1 charges of up to 3 kW.

The second source presents an average daily charging load per user for data obtained from December 2018 to January 2020 in a large housing cooperative in Norway [17]. The average daily load reported is plotted in Fig. 10. Based on the magnitude of this charging load, it can be inferred that these were small residential chargers as well. Nonetheless, the charging behavior and patterns recognized from [16] and [17] are resemblant of those estimated in this paper for methods 1 and 2.

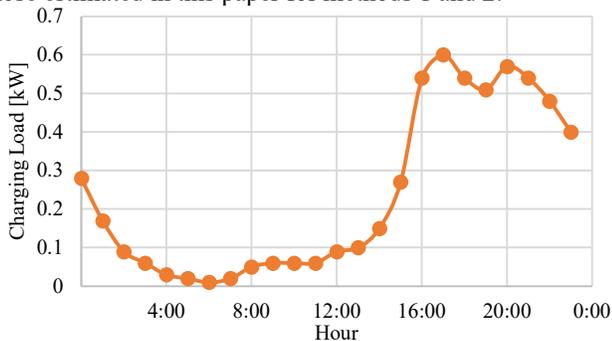

Fig. 10: Average EV charging load for residential case in Norway [17].

With charging behavior validation from these previous sources, our own generated datasets for charging times and load can be considered adequate datasets for level-2 residential EV charging for a US region, in particular the Texas coastal region. Our datasets can be used for power systems planning and operations for US utilities to understand what the new load demand can be depending on the expected number of EVs to be adopted at their service area, and assumed to follow the typical daily charging loads estimated in this work.

## VI. DATA AVAILABILITY

Two groups of data, generated with the methods presented in Sections II and III, pertaining to daily starting and ending charge times, and energy drawn during each charging event, are available in an open-access data repository [18], as well as the code scripts to load them in both Python and MATLAB.

## VII. CONCLUSIONS

Two different data-driven methodologies are presented to estimate residential electric vehicle (EV) charging load profiles, aiming to address the challenge of data scarcity from privacy-limited residential EV charging data. The methods were developed and applied using real-world data from CenterPoint Energy (CNP) customers in the greater Houston area. The first method used a least-squares state estimation approach to determine average EV charging power, and modeled starting and ending charge times based on statistical information derived from EV and non-EV meters load data differences. This method can be primarily useful for network upgrades planning by investigating the aggregate potential load increase based on expected EV penetration in a service area. The second method isolated the EV-specific load from the load difference between EV and non-EV meters, and used a kernel density estimation (KDE) to generate a data-driven probabilistic model for Monte Carlo simulations. This method is more useful for more immediate operations planning, such as day-ahead forecasting, with a smoother estimate of EV charging levels.

Both approaches consistently identified a distinct "u-shaped" aggregate daily EV charging load profile. Thus, key findings can be summarized as the EV charging activity being highly concentrated in the evening and late-night hours, with start times primarily clustering between 18:00 and 24:00 hours. This behavior results in a significantly prolonged peak at night and a clear load nadir during daytime hours. These charging patterns were validated by comparison with two other publicly available studies done with real-world data from UK and Norway, which reported similar trends.

The methodologies and estimated profiles presented in this paper provide a valuable and scalable way for utilities to plan for the potential grid impacts of widespread EV adoption. These data-driven models offer a practical way to forecast potential load increases on distribution networks with different levels of EV penetration.

## ACKNOWLEDGMENT

This work is part of a collaboration between the University of Houston and CenterPoint Energy (CNP), conducted under support by U.S. DOE Digitizing Utilities (Round 2) program.